\documentclass[preprint,secnumarabic,showpacs,amssymb, nobibnotes ,aps, prd]{revtex4-1}

\usepackage{graphicx}\usepackage{dcolumn}\usepackage{bm}
\begin{document}
\title{Impulsive gravitational waves of massless particles in extended theories of gravity}
\author{Morteza Mohseni}
\email{m-mohseni@pnu.ac.ir}
\affiliation{Physics Department, Payame Noor University, 19395-3697 Tehran, Iran}
\date{\today}
\begin{abstract}
We investigate the vacuum pp-wave and Aichelburg-Sexl-type solutions in $f(R)$ and the modified Gauss-Bonnet theories of gravity with both minimal 
and nonminimal couplings between matter and geometry. In each case, we obtain the necessary condition for the theory to admit the solution and examine 
it for several specific models. We show that the wave profiles are the same or proportional to the general relativistic one.   
\end{abstract}
\maketitle
\section{Introduction}
Extensions of the general theory of relativity in which functions of some geometric quantity are coupled either minimally or nonminimally to the matter 
part of the usual action for gravity have been widely studied recently, mainly as part of current efforts to explain the Universe with its observed 
late-time accelerated expansion \cite{smi,per,rei}. The so-called $f(R)$ gravity comprises the largest subset of models constructed in this way, see 
Refs. \cite{capo,faro,def} for reviews. In these models, a priori arbitrary functions of the scalar curvature of spacetime are included in the gravitational action
in its various metric, Palatini, or metric-affine formulations. The $f(R)$ function can also be coupled nonminimally to the matter Lagrangian, as  
suggested in Refs \cite{odi,noji}. An interesting consequence of such nonminimal coupling is the emergence of extra forces making the trajectories 
of otherwise-free particles nongeodesic \cite{ber}. Another well-known class consists of the so-called ${\mathcal F}({\mathcal G})$ gravity models, in 
which a function of the Gauss-Bonnet invariant, ${\mathcal G}$, is added to the Einstein-Hilbert action \cite{gau}. More general modified Gauss-Bonnet 
theories of gravity with nonminimal coupling have also been suggested \cite{gem1,gem2}, see also Ref. \cite{mor}.  

Various aspects of the above-mentioned models have been studied in recent years, including black hole solutions and their thermodynamics and 
cosmological solutions with accelerated expansion. There is now a huge literature on this, and a selection is listed in Refs. \cite{capo,faro,def} and 
\cite{odin}. Massless and massive gravitational wave solutions have been presented, too \cite{cpi}. The issue of linearized $f(R)$ gravity has been studied 
in Refs. \cite{lin1,lin2,lin3}. Compared with general relativity, several new features arise in this context, namely, the appearance of extra polarization modes and 
a nonlinear dispersion law, see, e.g., Ref. \cite{cord}.   

Different models introduced in this context have been examined against several theoretical or observational criteria. As an example, the study by Dolgov and 
Kawasaki \cite{dol} has shown that certain $f(R)$ models accommodate ghosts and hence are ruled out by the consequent instability. One can also mention
constraints coming from solar system effects such as planetary orbits or bending of light \cite{sol}, and the bounds coming from imposing the energy conditions 
\cite{santo}, in this regard.

The aim of the present work is to find classes of these extended theories of gravity which admit plane-fronted-parallel rays gravitational wave solutions the same or 
similar to the general relativistic vacuum or Aichelburg-Sexl solutions \cite{aic}. One motivation behind this work is the recent interest in gravitational pp-waves in general, 
and in gravitational shock waves of various sources, in particular. In fact, in the context of general relativity, the problem has been studied by boosting the Kerr metric 
in Refs. \cite{pend} and \cite{texa}, by boosting the Kerr-Newman metric in Ref. \cite{geo}, in the presence of a nonvanishing cosmological constant  \cite{hota}, for 
motion in Schwarzschild-Nordstr\"{o}m and Schwarzschild-de Sitter spacetimes  \cite{sfet}, for particles with arbitrary multipoles in Ref. \cite{pod}, for motion in 
Nariai universe \cite{nar}, and for motion in the presence of electromagnetic fields in Ref. \cite{orte}. A reason for interest in such solutions lies in their 
role in the scattering of particles off of each other at ultrahigh energies \cite{hoft,hoga}. Extensions of the Aichelburg-Sexl solution outside the context of 
general relativity have also been of interest, namely, in the framework of the brane-induced gravity \cite{kalo} and in the context of the ghost-free model 
of massive gravity \cite{moh}. 

Second, and at the same time, this study aims to put forward a new criterion for testing extended gravity models. The basic idea is that the gravitational 
radiation by moving particles, massless ones in the present case, is detectable, at least in principle, and whether a given model admits the plane wave 
solution distinguishes it from other models. For models admitting the solution, the explicit form of the wave profile provides further information.       

In the next sections, after a brief review of the Aichelburg-Sexl solution of general relativity,  we apply the above ideas to several well-known 
extended theories of gravity  including minimal and nonminimal $f(R)$ gravities and ${\mathcal F}({\mathcal G})$ gravity.
\section{The Aichelburg-Sexl solution}
The Aichelburg-Sexl (AS) solution, first introduced in Ref. \cite{aic}, represents the gravitational field of a massless particle moving in an 
otherwise empty space. It belongs to the general class of plane-fronted gravitational waves with parallel rays, or pp-wave, solutions.
The general form of a pp-wave line element may be written as
\begin{equation}\label{4a}
ds^2=-dudv-K(u,x,y)du^2+dx^2+dy^2,
\end{equation}
in which $u=t-z$ and $v=t+z$ are the null coordinates, and $K(u,x,y)$ is obtained from the field equations. For vacuum, the field
equation results in the two-dimensional Laplace equation to be satisfied by $K(u,x,y)$.   
   
The AS solution may be obtained by inserting the energy-momentum tensor of a massless particle in the Einstein 
equation. Starting with the action for the coupled  gravity-massless particle system,
\begin{equation}\label{l1}
S=\frac{1}{16\pi G}\int{\sqrt{-g}}\,R\,d^4x+\int{\sqrt{-g}}\,{\mathcal L}_{p}\,d^4x,
\end{equation}
one obtains
\begin{equation}\label{e1}
G_{\mu\nu}=8\pi GT_{\mu\nu},
\end{equation}
where 
\begin{eqnarray*}
R=\delta^\alpha_\mu g^{\beta\nu}(\partial_\alpha\Gamma^\mu_{\beta\nu}-\partial_\beta\Gamma^\mu_{\alpha\nu}+\Gamma^\mu_{\alpha\gamma}
\Gamma^\gamma_{\nu\beta}-\Gamma^\mu_{\beta\gamma}\Gamma^\gamma_{\nu\alpha})
\end{eqnarray*}
is the scalar curvature, $G_{\mu\nu}$ is the Einstein tensor,  and 
$$T_{\mu\nu}=-\frac{2}{\sqrt{-g}} \frac{\delta({\mathcal L}_{p}{\sqrt{-g}})}{\delta g^{\mu\nu}}$$ is the energy-momentum tensor. 
The above action is expressed in a system of units in which $c=1$. For convenience, we also set $8\pi G=1$. For a massless particle of momentum $p$ with
\begin{equation}\label{1c}
{\mathcal L}_{p}=\frac{p}{2}\int{g_{\mu\nu}{\dot x}^\mu{\dot x}^\nu}\delta^4(x-x(\tau))d\tau
\end{equation} 
in which ${\dot x}^\mu=\frac{dx^\mu}{d\tau}$, we have
\begin{equation}\label{2c}
T^{\mu\nu}=p\int\delta^4(x-x(\tau)){\dot x}^\mu{\dot x}^\nu d\tau,
\end{equation}
where $x^\mu(\tau)$ corresponds to the trajectory of the particle which has to be a null geodesic of the spacetime under consideration. Thus, we can take $x^\mu(\tau)$  to be of the form 
($\tau,0,0,z(\tau)=\tau)$ which is null and satisfies the geodesic equation in the spacetime described by Eq. (\ref{4a}). Inserting these together with the ansatz
given by Eq. (\ref{4a}) into Eq. (\ref{e1}) and making use of the following relation 
\begin{eqnarray*}
\left(\frac{\partial^2}{\partial x^2}+\frac{\partial^2}{\partial y^2}\right)\ln{\sqrt{x^2+y^2}}=2\pi\delta(x)\delta(y),
\end{eqnarray*}
we obtain 
\begin{equation}\label{t5}
K_{AS}(u,x,y)=-\frac{p}{\pi}\delta(u)\ln{\sqrt{x^2+y^2}}
\end{equation}
which upon substitution in Eq. (\ref{4a}), describes a gravitational shock-wave propagating with the speed of light along the $z$-direction.
\section{Gravitational waves in nonminimal $f(R)$ gravity}
We start with the following action \cite{ber}
\begin{equation}\label{1a}
S=\int{\sqrt{-g}}\left(\frac{1}{2}f(R)+(1+\lambda F(R)){\mathcal L}_m\right)d^4 x
\end{equation}
in which $f(R)$ and $F(R)$ are arbitrary functions of $R$. This represents an extension of the general theory of  relativity in which the matter field 
Lagrangian density ${\mathcal L}_m$ is coupled nonminimally with the geometric structure $F(R)$, and the coupling constant $\lambda$ controls
the strength of the coupling. This reduces to the usual $f(R)$ theories of gravity for $\lambda=0$, which in turn reduces to general relativity by 
choosing $f(R)=R$. 

The field equation associated with the above action reads
\begin{eqnarray}\label{1b}
{\mathcal E}_{\mu\nu}&=&\nabla_\mu\nabla_\nu f^\prime(R)-g_{\mu\nu}\square f^\prime(R)\nonumber\\&&+2\lambda(\nabla_\mu\nabla_\nu-
g_{\mu\nu}\square){\mathcal L}_m F^\prime(R)\nonumber\\&&-2\lambda F^\prime(R){\mathcal L}_m R_{\mu\nu}+(1+\lambda F(R))T_{\mu\nu},
\end{eqnarray}
where $\nabla$ means covariant differentiation, $\square=\nabla_\mu\nabla^\mu$, $R_{\mu\nu}$ is the Ricci tensor, and 
${\mathcal E}_{\mu\nu}=f^\prime(R)R_{\mu\nu}-\frac{1}{2}f(R)g_{\mu\nu}.$
 
Now, let us examine the above field equation to see if  the vacuum pp-wave solution is admitted in this model. For vacuum, the last three terms in the 
right-hand side of Eq. (\ref{1b}) vanish, and if we insert the ansatz  (\ref{4a}) into the resulting equation, we will reach the following relations
\begin{equation}\label{j72}
f^\prime(0)\left(\frac{\partial^2}{\partial x^2}+\frac{\partial^2}{\partial y^2}\right)K(u,x,y)-f(0)K(u,x,y)=0,
\end{equation} 
and
\begin{equation}\label{c1}
f(0)=0.
\end{equation}
Thus, the vacuum pp-wave solution is admitted only if the above condition holds. This condition is in fact among the requirements needed to constrain 
the models from cosmological or solar system tests, which is also consistent with the limiting case of the $\Lambda$CDM phenomenology \cite{saw}. 
The other requirement is that $f(R)$ should tend to a constant when the scalar curvature tends to infinity. 

When Eq. (\ref{c1}) holds, Eq. (\ref{j72}) reduces to the two-dimensional Laplace equation in the transverse plane, provided $f^\prime(0)\neq 0$.
This means that the waveform is the same as the vacuum general relativistic one. For the special case where $f^\prime(0)=0$, Eq. (\ref{j72}) is trivially 
satisfied with arbitrary $K(u,x,y)$. But this corresponds to the absence of the linear term $R$ in the Lagrangian and hence is ruled out. 

Now, we consider a massless particle moving along the $z$-direction and seek solutions of the form given in Eq.~(\ref{4a}) (renaming  $K(u,x,y)$ to 
$ K_{nR}(u,x,y)$ to avoid confusion). Inserting Eqs. (\ref{1c}) and (\ref{2c}) into the above field equation (with ${\mathcal L}_m$ replaced by 
${\mathcal L}_{p}$) and noting that for massless particles $g_{\mu\nu}{\dot x}^\mu{\dot x}^\nu$ vanishes, we reach  
\begin{equation}\label{j72c}
\nabla^2_TK_{nR}(u,x,y)=-2p\frac{1+\lambda F(0)}{f^\prime(0)}\delta(u)\delta(x)\delta(y),
\end{equation} 
provided Eq. (\ref{c1}) holds and $f^\prime(0)\neq 0$. Here, $\nabla^2_T\equiv\frac{\partial^2}{\partial x^2}+\frac{\partial^2}{\partial y^2}$. 
Thus, if Eq. (\ref{4a}) is satisfied, the pp-wave solution (\ref{4a}) is admitted, and the waveform is given by 
\begin{equation}\label{a71}
K_{nR}(u,x,y)=\frac{1+\lambda F(0)}{f^\prime(0)}K_{AS}(u,x,y).
\end{equation} 
If we also choose $F(0)=0$, then the effect of the nonminimal coupling disappears totally. For, $F(0)\neq 0$ the nonminimal coupling has a contribution
equal to $\frac{F(0)}{f^\prime(0)}\lambda$. Because of the very small expected value of the coupling constant $\lambda$, this would be a small contribution; 
see Ref. \cite{aron} for a discussion of bounds on the values of $\lambda$.

It should be noted here that for the pp-wave spacetime  (\ref{4a}), the nonlinear field equation (\ref{1b}) reduces 
to a linear equation discussed above, and this allows the use of distributional expressions as in general relativity. 
\section{Gravitational waves in $f(R)$ gravity}
The well-studied $f(R)$ theories of gravity are in fact a subclass of the nonminimal theory considered above, with $\lambda=0$. 
Thus, both the vacuum pp-wave and the AS solutions are admitted if Eq. (\ref{c1}) holds.

To determine the waveform of the AS solution, we insert $\lambda=0$ into Eq. (\ref{a71}) (this time with $K_R(u,x,y)$ in place of $K_{nR}(u,x,y)$).
This yields 
\begin{equation}\label{a72}
K_R(u,x,y)=\frac{1}{f^\prime(0)}K_{AS}(u,x,y).
\end{equation} 
Since in general  $f^\prime(R) > 0$, otherwise ghosts are allowed;  the signs of $K(u,x,y)$ and $K_{AS}(u,x,y)$ are the same. 

An example of the models satisfying the requirement given by Eq. ({\ref{c1}) is the broken power-law model
$$f(R)=R-m^2\frac{c_1\left(\frac{R}{m^2}\right)^n}{1+c_2\left(\frac{R}{m^2}\right)^n}$$
suggested in Ref. \cite{saw}, in which $c_1, c_2, m$, and $n$ are constants, and $n > 0$. Note that the linear term above is included to reproduce the 
Einstein-Hilbert action in Eq. (\ref{1a}). For this model, we have 
\begin{eqnarray*}
\lim_{R\to 0}\frac{1}{f^\prime(R)}=\left\{\begin{array}{cll}
1 &  \mbox{if} & n>1\\  \frac{1}{1-c_1} & \mbox{if} & n=1 .\\ 0 & \mbox{if} & n<1
\end{array}\right.
\end{eqnarray*}
By inserting this into Eq. (\ref{a72}), we conclude that for $n\geq 1$, the model admits the plane wave solution, 
with a wave profile the same as the general relativistic one for $n>1$, and $\frac{1}{1-c_1}$ times the general relativistic waveform for $n=1$. 
For $n < 1$, in which the above limit equals zero, the solution reduces to the Minkowski spacetime. In other words, for $n<1$, the plane wave 
solution is not admitted, and this is in agreement with what we expect from the general requirement $f^\prime(R)>0$ mentioned earlier. 

Another example is the Starobinsky model \cite{star} given by $$f(R)=R+\frac{R^2}{M^2}.$$ Here, we have $$\frac{1}{f^\prime(0)}=1,$$
and hence the solution is exactly the same as the general relativistic one. 

Also, the following cosmologically viable model proposed in Ref. \cite{stars} (see also Ref. \cite{amen})
$$f(R)=R-\lambda_0 R_0\left(1-\frac{1}{\left(1+\frac{R^2}{R^2_0}\right)^n}\right),$$
in which $\lambda_0, R_0, n$ are positive constants, satisfies the condition (\ref{c1}). For this model, we have $f^\prime(0)=1$, i.e. coincidence with general 
relativity, again. However, because in this model $f^{\prime\prime}(0) < 0$, it is unstable \cite{hyu}. 

The model described by \cite{appl} 
$$f(R)=R+\frac{1}{a}\ln(\cosh(aR)+b\sinh(aR)),$$ where $a, b$ are constants, admits the solution, too. Here, we have $$\frac{1}{f^\prime(0)}=\frac{1}{1+b}.$$

An example of the models incompatible with the condition (\ref{c1}) is the so-called IR-modified gravity model of Refs. \cite{cap,car} described by 
$$f(R)=R-\frac{\mu^4}{R},$$ where $\mu\sim H_0$, $H_0$ being the Hubble constant. However, such models are ruled out by the Dolgov-Kawasaki instability \cite{dol}. 
It is interesting to note that in fact the above model and the Starobinsky model (the second example model discussed above) are special cases of 
a more general model given by $$f(R)=R-(1-n)\mu^2\left(\frac{R}{\mu^2}\right)^n,$$ which have been investigated in Ref. \cite{fau}.

Another model which does not admit the solution is given by \cite{nju} $$f(R)=R+\alpha\ln\left(\frac{R}{\mu^2}\right)+\beta R^m, $$ $\alpha,\beta, m$ being constants.
\section{Gravitational waves in nonminimal ${\mathcal F}({\mathcal G})$ gravity}
In this section, we consider a nonminimal ${\mathcal F}({\mathcal G})$ gravity model described by the following action \cite{mor}
\begin{equation}\label{1aa}
S=\int{\sqrt{-g}}\left(\frac{1}{2}R+{\mathcal F}({\mathcal G})+(1+\kappa{\mathcal H}({\mathcal G})){\mathcal L}_m\right)d^4 x,
\end{equation}
in which an arbitrary function of the Gauss-Bonnet invariant 
\begin{equation}\label{I2}
{\mathcal G}\equiv R^2-4R_{\mu\nu}R^{\mu\nu}+R_{\mu\nu\lambda\sigma}R^{\mu\nu\lambda\sigma}
\end{equation}
is coupled to the matter field Lagrangian ${\mathcal L}_m$ with a coupling constant 
$\kappa$. This represents a generalized version of the actions introduced in Refs. \cite{gem1,gem2}. 

The equation of motion resulting from the above action  is given by
\begin{eqnarray}\label{h13}
(1+\kappa{\mathcal H(\mathcal G)})T^{\mu\nu}&=&G^{\mu\nu}-g^{\mu\nu}{\mathcal F}({\mathcal G})\nonumber\\&&
+4H^{\mu\nu}({\mathcal F}^\prime({\mathcal G})+\kappa{\mathcal L}_m{\mathcal H}^\prime({\mathcal G})),
\end{eqnarray}
where $${\mathcal F}^\prime({\mathcal G})=\frac{d{\mathcal F}({\mathcal G})}{d{\mathcal G}},$$ and 
\begin{eqnarray*}
H^{\mu\nu}&=&RR^{\mu\nu}+{R^\mu}_{\alpha\beta\gamma}R^{\nu\alpha\beta\gamma}-2R^\mu_\alpha 
R^{\alpha\nu}\\&&+2R^{\mu\alpha\beta\nu}R_{\alpha\beta}-2G^{\mu\nu}\nabla^2-
R\nabla^\mu\nabla^\nu\\&&-2g^{\mu\nu}R^{\alpha\beta}\nabla_\alpha\nabla_\beta+2R^{\alpha\nu}\nabla_\alpha\nabla
^\mu\\&&+2R^{\mu\alpha}\nabla_\alpha\nabla^\nu-2R^{\mu\alpha\beta\nu}\nabla_\alpha\nabla_\beta.
\end{eqnarray*}
Now, we consider the pp-wave anstaz, Eq. (\ref{4a}). First, we note that the Gauss-Bonnet invariant, Eq. (\ref{I2}) above, vanishes identically
for the spacetime described by Eq. (\ref{4a}). For vacuum, the above field equation reduces to
\begin{equation}\label{j72a}
\left(\frac{\partial^2}{\partial x^2}+\frac{\partial^2}{\partial y^2}\right)K(u,x,y)-2{\mathcal F}(0)K(u,x,y)=0,
\end{equation} 
and
\begin{equation}\label{6d} 
{\mathcal F}(0)=0.
\end{equation}
Thus, the vacuum solution is admitted only if the above condition holds. Then, Eq. (\ref{j72a}) reduces to the same equation governing the 
waveform in general relativity. Hence, the waveform is the same as the general relativistic counterpart.

Back to the nonminimal coupling to a massless particle, by taking ${\mathcal L}_{p}$ as the matter Lagrangian and inserting the associated 
energy-momentum tensor into the field equation (\ref{h13}), we obtain the plane wave solution (\ref{4a}) whenever  Eq. (\ref{6d}) is satisfied. 
The relevant waveform  is given by
\begin{equation}\label{z1}
K_{\mathcal G}(u,x,y)=(1+\kappa{\mathcal H}(0))K_{AS}(u,x,y).
\end{equation}
If, in addition, the function ${\mathcal H}({\mathcal G})$ is chosen so that it satisfies ${\mathcal H}(0)=0$, then the waveform is not 
distinguishable from the general relativistic one.
\section{Gravitational waves in ${\mathcal F}({\mathcal G})$ gravity}
The action for ${\mathcal F}(\mathcal G)$ gravity is given by 
\begin {equation}\label{1.1}
S=\int d^{4}x\sqrt{-g}\left[\frac{1}{2}R+{\mathcal F}({\mathcal G})+{\mathcal L}_m\right]\,, 
\end{equation}
which is a particular case of  the nonminimal model discussed in the previous section with $\kappa=0$. Thus, all such models satisfying Eq. (\ref{6d})
admit both the vacuum pp-wave solution and the AS solution. Now, from Eq. (\ref{z1}), with vanishing $\kappa$, it is obvious that the waveform 
$K_{\mathcal G}(u,x,y)$ is the same as the one in general relativity.

Examples of the ${\mathcal F}({\mathcal G})$ models satisfying the requirement Eq. (\ref{6d}) include 
$${\mathcal F}({\mathcal G})={\mathcal G}^n,$$ with $n > 0$ which is shown in Ref. \cite {dave} that it could also pass solar system tests 
for $n\lesssim 0.074$.   

Also, the following cosmologically viable models
\begin{eqnarray}
{\mathcal F}({\mathcal G})&=&\mu\frac{\mathcal G}{{\mathcal G}_\star}\tan^{-1}\left(\frac{\mathcal G}{{\mathcal G}_\star}\right)-\frac{\mu}{2}{\sqrt{{\mathcal 
G}_\star}}\ln\left(1+\frac{{\mathcal G}^2}{{{\mathcal G}_\star}^2}\right)\nonumber\\&&-\alpha\mu{\sqrt{\mathcal G}_\star},\label{14a}\\
{\mathcal F}({\mathcal G})&=&\mu\frac{\mathcal G}{{\mathcal G}_\star}\tan^{-1}\left(\frac{\mathcal G}{{\mathcal G}_\star}\right)-\alpha\mu
{\sqrt{\mathcal G}_\star},\label{14b}\\
{\mathcal F}({\mathcal G})&=&\mu{\sqrt{{\mathcal G}_\star}}\ln\cosh\left(\frac{{\mathcal G}^2}{{{\mathcal G}_\star}^2}\right)-\alpha\mu
{\sqrt{\mathcal G}_\star},\label{14c}
\end{eqnarray}
proposed in Ref. \cite{feli} admit the plane wave solution for $\alpha=0$. Here, $\mu$ and ${\mathcal G}_\star$ are positive constants.

The model presented in Ref. \cite{nori} provides another example which admits the wave solution. It is described by
\begin{eqnarray*}
{\mathcal F}({\mathcal G})=\frac{({\mathcal G}-G_0)^{2n+1}+G^{2n+1}_0}{F_0+F_1\{({\mathcal G}-G_0)^{2n+1}+G^{2n+1}_0\}},
\end{eqnarray*}
where $F_0, F_1, G_0$ are constants.

It is interesting to note that both $f(R)$ and ${\mathcal F}({\mathcal G})$ theories considered above can be obtained from a more general theory 
described by the following action \cite{eliz}
\begin{eqnarray}\label{1.2}
S=\int{\sqrt{-g}}(F(R,{\mathcal G})+{\mathcal L}_m)d^4x,
\end{eqnarray}
which can be seen from the associated field equation which admits the above plane wave solution when similar conditions to those introduced above 
are satisfied.
\section{Discussion and conclusions}\label{phys}
In this work, we studied plane-fronted gravitational waves with parallel rays in the context of some extended theories of gravity. We considered
$f(R)$ and modified Gauss-Bonnet gravity with minimal and nonminimal couplings to matter and showed that they admit the vacuum pp-wave solution
and also an Aichelburg-Sexl-type solution  if certain conditions are satisfied. For $f(R)$ gravity, the required condition is that $f(R)$ vanishes 
for vanishing scalar curvature. This condition is compatible with the requirements for such theories to pass local gravity tests. It is also 
the same condition for a given model to admit the Schwarzschild solution. Thus, for those models admitting the Schwarzschild  solution, it should be possible
to obtain the AS plane wave solution by boosting the black hole one as in general relativity. The explicit form of the wave profile depends on $f^\prime(0)$ and 
coincides with the general relativistic wave profile for some specific models including the well-known Starobinsky model. A similar condition holds for the 
modified Gauss-Bonnet gravity. In the latter case, the solution is the same as the general relativistic one. This was examined for several specific 
cosmologically viable models. For models with nonminimal coupling between the matter and geometry, more interesting options are available, 
including the possibility of (dis)appearance of (the)a contribution from the nonminimal coupling by choosing appropriate coupled function. 

The gravitational wave solution presented here might be used as an experimentally testable, at least in principle, criterion to distinguish between various extended 
gravity models. This can be achieved by looking at the behavior of two pointlike objects in the gravitational field of the massless source. For models with minimal 
coupling, this can be seen by measuring the relative acceleration of two nearby test particles separated by $n^\mu$ which is obtained 
from the geodesic deviation equation
\begin{equation}\label{z2}
\frac{D^2n^\mu}{D\tau^2}=-{R^\mu}_{\alpha\nu\beta}{\dot x}^\alpha n^\nu{\dot x}^\beta
\end{equation}
and noting that for the spacetime under consideration, the components of the Riemann curvature tensor are proportional to the second derivatives of the wave 
profile $K(u,x,y)$ with respect to the transverse coordinates. For models with nonminimal couplings where the particles do not move along geodesics 
as a result of extra forces coming from the coupling, the above equation should be modified by adding the relevant terms. However, since the extra force is 
proportional to the gradient of the scalar curvature or the Gauss-Bonnet invariant, both of which vanish in the above considered 
spacetime, the same argument still holds. The fact that the waveform obtained by application of the extended theories of gravity is proportional to the general
relativistic waveform would then reflects in the observed relative accelerations. 

The results obtained here might also be used to study the scattering of particles at high energies in the framework of extended theories of gravity. Possible interesting extensions of the 
present work include a study of shock waves due to massless particles moving in curved
backgrounds, and particles with arbitrary multipoles moving in curved spacetimes in the presence of matter fields and-/or a cosmological constant.
\section*{Acknowledgments}
The author gratefully acknowledges several comments by two anonymous referees of Physical Review D. 

\end{document}